\documentclass[12pt,a4paper]{article}
\usepackage{epsf,epsfig,wrapfig}
\newcommand{\dirac}{\displaystyle{\not}}

\begin{document}
\begin{center}
\textbf{DOMINANT CONTRIBUTION IN PION PRODUCTION\\ SINGLE-SPIN ASYMMETRIES}\footnote{Talk presented at the XII Advanced Research Workshop on High Energy Spin Physics (DSPIN--07), Dubna, 3--7 Sept.~2007.}

\vspace{5mm}
Philip G. Ratcliffe$^{1,2\,\dag}$, \underline{Marco Ramilli}$^{1}$

\vspace{5mm}
{\small (1)
\emph{Dipartimento di Fisica e Matematica, Universit\`{a} degli Studi dell'Insubria}
}
\\
{\small
(2) \emph{Istituto Nazionale di Fisica Nucleare, Sezione di Milano}
\\
$\dag$ \emph{E-mail: philip.ratcliffe@uninsubria.it}
}
\end{center}

\vspace{5mm}
\begin{abstract}
  Working with a completely collinear twist-3 factorized cross-section formula,
  we identify two largely dominant partonic sub-processes, which contribute to
  the single-spin asymmetries in semi-inclusive pion production, in the region
  of large $p_T$ and medium--large $x_F$.
\end{abstract}
\section{Introduction}
During the past years, different models have been developed in an attempt to
explain the mechanism behind the single-spin asymmetries observed experimentally
in high-energy hadronic interactions. The approach based on the study of the
hadronic cross-section contribution given by the twist-3 components in the
operator product expansion of parton matrix elements turns out to be
particularly interesting: taking into account such terms provides a consistent
model. However, at the same time the complexity of the calculational framework
unfortunately increases, since twist-3 contribution are characterized by the
presence of an additional gauge-field term, which in turn implies an \emph{extra
gluon} in the sub-processes, see for example~\cite{ET,QiuSterman98}.

Restricting our analysis therefore to a particular class of processes (pion
production in proton--proton collisions), our principal aim is to identify
which, if any, among all possible partonic sub-processes provide the dominant
contributions to the asymmetry and to understand the origin of the suppression
of the other terms. We can thus list a set of criteria (which we call
\emph{``selection rules''}) summarizing these mechanisms. To simplify our
analysis, we shall extract a totally collinear cross-section formula, in the
axial gauge and in the limit of $x_F\to1$, valid for large $p_T$.
\section{The model}
We shall now go into detail, first by providing an expression for the
twist-3 contribution to the cross-section through the study of the pole behavior
of the Bjorken variables, and then by analyzing the causes of the suppression of
many other sub-processes.
\subsection{The poles}
Working in axial gauge, thus setting $A^+=0$, allows us to write the twist-3
contribution to the cross-section in the following way:
\begin{equation}
d\sigma^{(\tau=3)} \simeq
\mbox{Tr}\left\{\Phi_A^\alpha(x_1,x_2)\mbox{S}^\beta(x_1,x_2)\right\}
g_{\perp\alpha\beta},
\end{equation}
where $\Phi_A^\alpha(x_1,x_2)$ is the multi-parton matrix element and the index
$\alpha$ is completely transverse, due to gauge choice. Moreover, in the axial
gauge, the relation between $\Phi_A^\alpha(x_1,x_2)$ and
$\Phi_F^\alpha(x_1,x_2)$ assumes a very simple form (see \cite{physics-report},
Eq.~7.3.30):
\begin{equation}
(x_2-x_1)\Phi_A^\alpha(x_1,x_2)=-i\Phi_F^\alpha(x_1,x_2),
\end{equation}
demonstrating that if $\Phi_F^\alpha(x_1,x_2)$ is different from zero for
$x_1=x_2$, then $\Phi_A^\alpha(x_1,x_2)$ must have a pole.

The analysis of the hard part is also crucial for the pole structure; there are
two different possibilities for the extra gluon, generated at twist-3, to
interact significantly: with the on-shell fragmenting parton (the so-called
final-state interactions, FSI) and with the on-shell parton coming from the
unpolarized nucleon (initial-state interactions, ISI); the important feature of
these interactions is the presence of an extra internal propagator, whose Dirac
structure has the form
\begin{equation}\label{prop}
  \cdots \frac{\dirac k}{2(P\cdot
  k)}\left(\frac{2k_\alpha-(x_2-x_1)\gamma_\alpha\dirac
  P}{x_2-x_1-i\varepsilon}\right)\cdots,
\end{equation}
where $k^\mu$ is the four-momentum of the on-shell parton and $P^\mu$ is the
four-momentum of the polarized hadron.

By also taking into account the pole behavior originating in the multi-parton
matrix element, it is possible to separate the trace over the Dirac indices into
two traces, each one with a different pole structure: the first, known as the
\emph{single-pole} contribution, where the $(x_2-x_1)$ term in the numerator
cancels the pole contribution of the matrix element, and the other, called the
\emph{double-pole} contribution, where no such cancelation occurs. In order to
maintain the cross-section a real quantity, we are forced to take the imaginary
part of these poles, remembering that
\begin{eqnarray}
  \mbox{Im}\left(\frac{1}{(x_2-x_1\pm i\varepsilon)}\right)& = &
  \mp i\pi \delta (x_2-x_1),\\
  \mbox{Im}\left(\frac{1}{(x_2-x_1\pm i\varepsilon)^2}\right)& = &
  \mp i\pi \delta^\prime(x_2-x_1).
\end{eqnarray}
Using these relations and integrating the derivative of the delta function by
parts, we obtain the following expression for the twist-3 contribution to the
cross-section:
\begin{eqnarray}
  d\sigma^{(\tau=3)} &=&
  \int dx\,dx^\prime \; \frac{dz}{z^3} \; \varepsilon_T^{P_hS_\perp}
  \left\{\frac{dG_F(x,x)}{dx} H_{DP}(x,x^\prime,z)\right.
  \nonumber\\
  && \hspace*{10em}\left.\vphantom{\frac{dG_F(x,x)}{dx}}
  + G_F(x,x)H_{SP}(x,x^\prime,z)\right\} f(x^\prime) \, D(z),
\end{eqnarray}
where we have omitted the color factors and the sum over flavor indices;
$\varepsilon_T^{\mu\nu}$ is the antisymmetric tensor in the transverse
directions, $G_F(x,x)$ is the multi-parton distribution function evaluated at
the pole (owing to the delta functions), $f(x^\prime)$ is the unpolarized quark
density and $H$ represents the hard-scattering partonic cross-sections, with
$DP$ and $SP$ standing respectively for double pole and single pole.

\subsection{``Selection Rules''}
Given such an expression for the cross-section at twist three, we list here the
set of principles we have adopted to identify the possibly dominant
contributions:
\begin{itemize}
\item[-]
  first, we expect DP contributions to be much more relevant than SP ones, owing
  to the presence of the derivative of the multiparton density function, which
  endows the asymmetry with a behavior in $x$ roughly as
  $A_N\sim\frac{1}{(1-x)}$ (for $x_F$ approaching unity, the Bjorken $x$ of the
  incoming parton also approaches unity), thus enhancing the contribution of
  such terms for growing $x_F$;
\item[-]
  for $x_F\to1$ and $|T|\ll{}|U|\ll{}|S|$, we expect the $t$-channel
  diagrams to be dominant; for the same reason, remembering the power
  suppression of the hard parts given in Eq.~\ref{prop}, we expect FSI
  to give a greater contribution than ISI;
\item[-]
  we neglected the contributions given by polarized gluons and by sea quarks
  since these may reasonably be expected to be small.
\end{itemize}

In order to test our model and the selection rules described above, we have
evaluated the single-spin asymmetries for the reaction $p^\uparrow p \to \pi^0
+X$ for the STAR kinematical range ($\sqrt{S}=200$\,GeV and
1.3\,GeV/$c<P_{hT}<2.8$\,GeV/$c$, see for example \cite{Nogach06}). Restricting
our analysis to the contribution given only by the $t$-channel diagram involved
in the process, in Fig.~1a we present a comparison between the data points and
the resulting prediction given by our model; we note that there is good
agreement with data for values of $x_F$ greater than $0.4-0.5$.

\begin{figure}[hbt]
  \begin{center}
    \begin{tabular}{cc}
      \mbox{\epsfig{figure=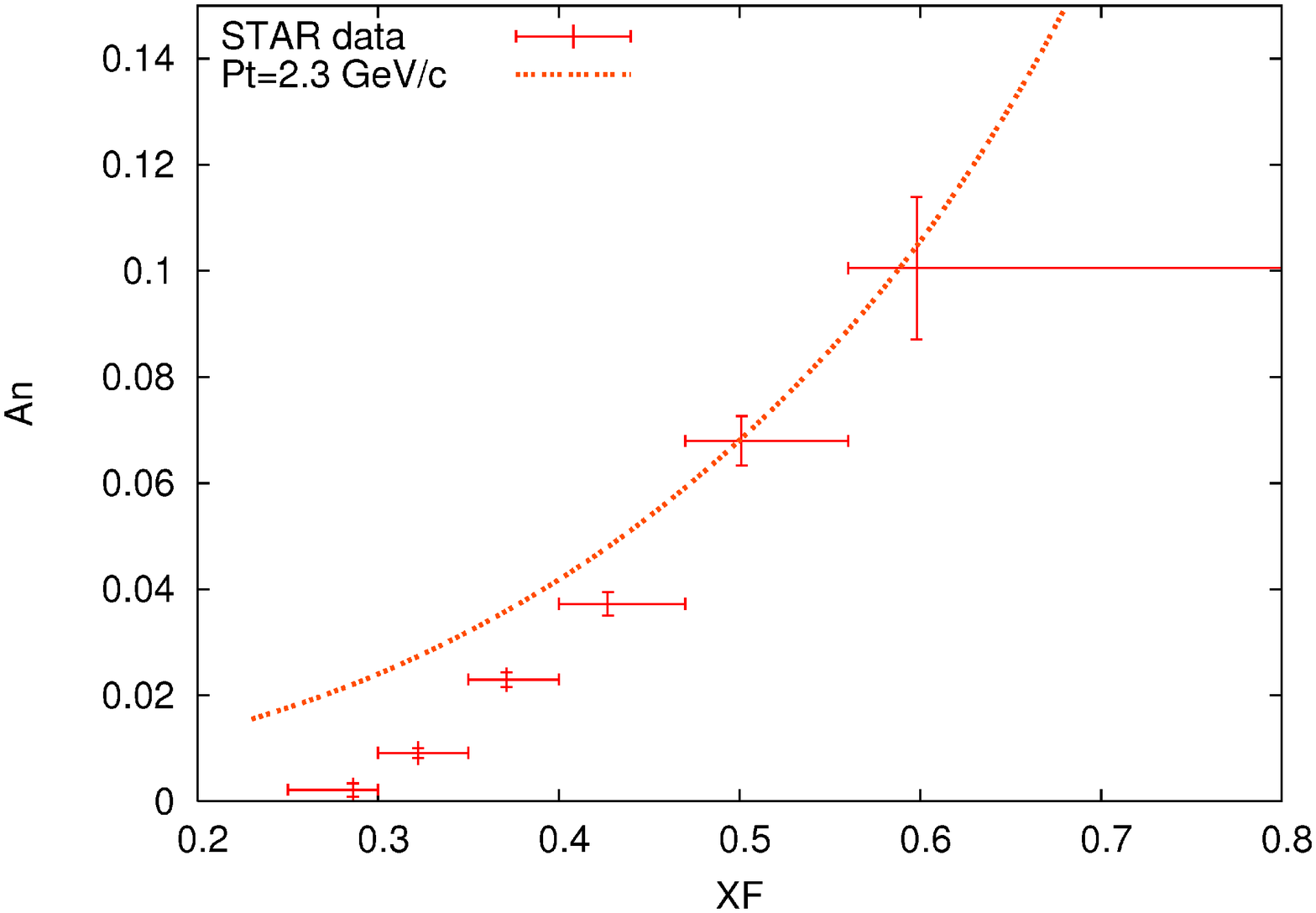,width=0.48\textwidth}} &
      \mbox{\epsfig{figure=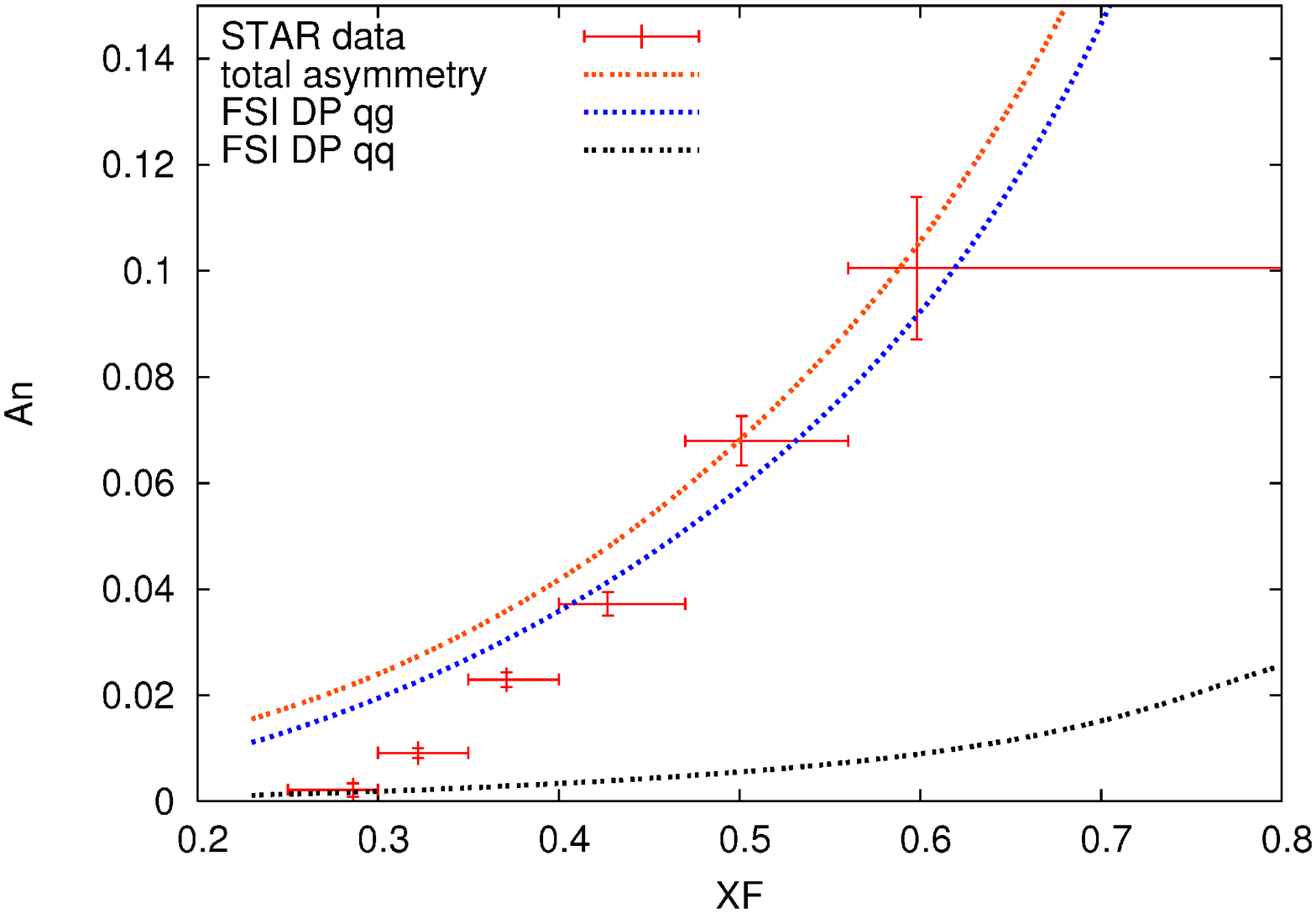, width=0.48\textwidth}}
    \\
      {\bf(a)}& {\bf(b)}
    \end{tabular}
  \end{center}
  {\small
  \textbf{Figure 1.}
  \textbf{(a)} The theoretical curve represents the prediction for the
  SSA in $\pi^0$ production evaluated at $P_{hT}=2.3$ GeV/c, compared to STAR
  data points.
  \textbf{(b)} Here we plot the same curve as in Fig.~1a, compared to the FSI
  DP term in a quark--gluon (here labeled \emph{qg}) sub-process and
  the FSI DP in a quark--quark subprocess.}\\
\end{figure}
In Fig.~1b we also plot the total asymmetry, but together with the contribution
given by the two major sub-processes we have identified, i.e.\ the $t$-channel
FSI DP terms. Comparing these curves, we can see how the two sub-processes
mentioned provide almost entirely the value of the asymmetry in the kinematical
range of $x_F>0.4$; for lower values of this variable, we expect all the
neglected contribution to become more important.
\section{Conclusions}
To summarize then:
\begin{itemize}
\item[-]
we have obtained an expression providing predictions for the single-spin
asymmetries for pion production consistent with data, in a completely collinear
framework, without appealing to any collinear expansion;
\item[-]
using such an expression and a simple set of criteria, we have also been
able to identify two largely dominant subprocesses, which are almost entirely
responsible for the asymmetries in the $x_F\to1$ limit.
\end{itemize}

%
\end{document}